\documentstyle[12pt,rei1]{article}
\begin{document}

%\pagestyle{myheadings}
%\markright{DRAFT: \today\hfill}

\title{Reheating of the Universe and Population III}
\author{Jeremiah P.\ Ostriker\altaffilmark{1} and 
Nickolay Y.\ Gnedin\altaffilmark{1,2}}
\altaffiltext{1}{Princeton University Observatory, Peyton Hall, 
Princeton, NJ 08544;\\ e-mail: \it jpo@astro.princeton.edu} 
\altaffiltext{2}{Department of Physics, Massachusetts Institute of Technology,
Cambridge, MA 02139; e-mail: \it gnedin@arcturus.mit.edu}

%$$
%  \framebox{$\displaystyle\phantom{\prod}$DRAFT: \today}
%$$

\load{\scriptsize}{\sc}

\def\A{{\cal A}}
\def\B{{\cal B}}
\def\ion#1#2{\rm #1\,\sc #2}
\def\HI{\ion{H}{i}}
\def\HII{\ion{H}{ii}}
\def\GI{\ion{He}{i}}
\def\GII{\ion{He}{ii}}
\def\GIII{\ion{He}{iii}}
\def\MH{{{\rm H}_2}}
\def\Hp{{{\rm H}_2^+}}
\def\Hm{{{\rm H}^-}}

\def\dim#1{\mbox{\,#1}}

\def\figdir{.}
\def\placefig#1{#1}

\def\capMB{{\it a)\/}
The Jeans mass for the adiabatically expanding intergalactic gas
({\it dashed line\,}), gas at $10^4\dim{K}$ ({\it dotted line\,}),
and transition curve for instantaneous reheating ({\it solid line\,}).
Dots show the nonlinear mass. {\it b)\/} The nonlinear mass 
({\it filled dots\,}) and the Jeans mass from
run A ({\it solid line\,}) and run B ({\it dashed line\,}).
}

\def\capZM{
Total mass $M_{\rm tot}$ ({\it upper left panel\,}), baryonic mass 
$M_{\rm b}$) ({\it upper right panel\,}), stellar mass $M_*$) ({\it lower 
left panel\,}), and the mass in metals $M_Z$) ({\it lower right\,}) as
a function of average redshift of star formation for all bound objects with
more than 100 particles at $z=4.3$. Open circles and filled squares show
Pop III and Pop II respectively.
}

\def\capZE{
Distribution of all bound objects with more than 100 particles identified at
$z=12.1$ in ($M_{\rm tot}$,$z_{\rm form}$) plane for three different 
redshifts: $z=12.1$ ({\it open circles\,}), $z=7.1$ ({\it stars\,}), and
$z=4.3$ ({\it filled squares\,}). Vertical lines with symbols mark 
correspondent redshifts. Note that two populations continue to exist at
both latter epochs but continued star formation depletes the pure
Pop III groups.
}

\def\capGF{
The star formation rate for run A ({\it solid lines\,}) and run B 
({\it dashed lines\,}) as a function of redshift. Bold lines show the
star formation rate as computed, and thin lines show the star formation rate
as would be computed if the cooling time were infinitely short. Thin dotted
line shows the rate with which the mass fraction of all halos with
the virial temperature above $1.5\times10^4\dim{K}$ increases as computed
from the Press-Schechter formalism.
}

\def\capTD{
Joint distribution for all star particles in temperature and
redshift at the moment and place of formation for run A. 
Two epochs of star formation:
high $z$ -- low $T$ and lower $z$ -- higher $T$ are clearly distinguishable.
}

\def\capLL{
The cooling function for the primeval gas with the $10^{-3}$ molecular
hydrogen abundance by mass ({\it heavy dots}). Shown with the thin line
is the value of the cooling function required for a virialized $3\sigma$
peak with a given temperature to cool in a Hubble time. Note distinct epochs
where $\MH$ band cooling or ($\HI$,$\GII$) line cooling permits collapse.
}

\def\tableone{
\begin{table}
\caption{Model and Numerical Parameters}
\medskip
$$
\begin{tabular}{cccccc}
Run & $N$ & Box size & Total mass res.\
 & Spatial res.\ & Dyn.\ range \\ \tableline
A & $128^3$ & $2h^{-1}{\rm\,Mpc}$ & $3.7\times10^5h^{-1}\dim{M}_{\sun}$ & $1.0h^{-1}{\rm\,kpc}$ & $2000$ \\
B & $ 64^3$ & $2h^{-1}{\rm\,Mpc}$ & $2.9\times10^6h^{-1}\dim{M}_{\sun}$ & $3.0h^{-1}{\rm\,kpc}$ & $640$ \\
C & $ 64^3$ & $1h^{-1}{\rm\,Mpc}$ & $3.7\times10^5h^{-1}\dim{M}_{\sun}$ & $1.5h^{-1}{\rm\,kpc}$ & $640$ \\
\end{tabular}
$$
\end{table}
}

\begin{abstract}

We note that current observational evidence strongly favors a conventional 
recombination of ionized matter subsequent to redshift $z=1200$,
followed by reionization prior to redshift $z=5$ and compute how this would
have occurred in a standard scenario for the growth of structure. Extending
prior semi-analytic work, we show by direct, high-resolution
numerical simulations (of a {\it COBE\/} normalized CDM+$\Lambda$ model)
that reheating, will occur in the interval $15>z>7$, followed by reionization
and accompanied by a significant increase in the Jeans mass. However,
the evolution of the Jeans mass does not significantly affect star formation
in dense, self-shielded clumps of gas, which are detached from the thermal
evolution of the rest of the universe. On average, the growth of the
Jeans mass tracks the growth of the nonlinear mass scale, a result we
suspect is due to nonlinear feedback effects.
Cooling on molecular
hydrogen leads to a burst of star formation prior to
reheating
which produces Population III stars with $\Omega_*$ reaching $10^{-5.5}$
and $\bar Z/Z_{\sun}$ reaching $10^{-3.7}$ by $z=14$.
Star formation subsequently slows down
as molecular hydrogen is depleted by photo-destruction and the rise of the
temperature. At later times, $z<10$, when the characteristic
virial temperature of gas clumps reach $10^4$ degrees, star formation 
increases again as hydrogen line cooling become efficient.
Objects containing Pop III stars accrete mass with time and, as soon
as they reach $10^4\dim{K}$ virial temperature, they engage in renewed
star formation and turn into normal Pop II objects having an old
Pop III metal poor component.

\end{abstract}


\section{Introduction}

In the standard, hot big bang model of cosmology, radiation and matter 
decouple from one another at approximately redshift $z\approx1200$ with
recombination proceeding until the ionized fraction freezes out
at $n_e/n_{\HI}\approx4\times10^{-4}$ at about redshift $z\approx800$
(cf.\ Peebles\markcite{P93} 1993). But we know from the classic 
Gunn-Peterson test (Gunn \& Peterson\markcite{GP65} 1965), 
that the universe is again ionized at low redshifts with the neutral fraction
still below $6\times10^{-6}$ at redshift $z\approx4$ 
(Jenkins \& Ostriker\markcite{JO91} 
1991; Giallongo et al.\markcite{Gea94} 1994). In this standard picture
reionization must occur, due to unknown sources in the intervening time 
interval. 
The purpose of the present paper is to indicate how and when this
happens according to present theories for the development of structure.
This work differs from prior papers on this subject primarily in 
two respects: (1) we treat (imperfectly but) in greater detail than has been
done previously many of the relevant physical processes such as molecular
hydrogen formation, spectrum of the ambient radiation field, absorption
etc, and (2) we allow explicitly for the density inhomogeneities that
develop through the action of gravity and pressure using the SLH code
of Gnedin\markcite{G95} (1995) as modified by Gnedin \& 
Bertschinger\markcite{GB96} (1996).

Many authors have studied this process, from the pioneering work of
Couchman \& Rees\markcite{CR86} (1986), Shapiro\markcite{S86a,S86b} (1986a,
 1986b) to the recent careful investigations by
Tegmark and Silk\markcite{TS94,95} (1994, 1995), Kawasaki \&
Fukigita\markcite{KF94} (1994), Giroux \& Shapiro\markcite{GS} (1996),
Tegmark et al.\markcite{Tea96} (1996) and others.
While varying amounts of physically detailed modelling were included
in all of those papers, there was a critical factor that could not be
followed in any of the semi-analytic treatments: the clumping of the gas.
But all relevant processes are dependent on the clumping factor
$\eta\equiv\langle\rho^2\rangle/\langle\rho\rangle^2$. Included among these
processes are {\it recombination, cooling, gravitational collapse, molecular 
hydrogen formation\/} and numerous others. Thus a fully nonlinear three 
dimensional threatment appears to be warranted. That is what we have
undertaken, using one of the currently plausible models for the growth
of structure - the CDM+$\Lambda$ model (cf.\ Ostriker \& 
Steinhardt\markcite{OS95} 1995 for references) as a point of departure.
The Jeans mass after decoupling for gas at the CBR temperature is
constant and has the value of $1.7\times10^6 h^{-1}{\rm\,M}_{\sun}$. After
redshift approximately 100, the gas temperature is no longer coupled to the
CBR temperature, and the Jeans mass decreases as $(1+z)^{3/2}$, approaching
the value of $5.4\times10^4 h^{-1}{\rm\,M}_{\sun}$ at $z=10$. So, it is
necessary to perform simulations with at least this mass resolution.
The work reported on here adopts the the model and numerical parameters given 
in Table 1 (close to the ``concordance'' model of Ostriker \& Steinhardt 
1995), but the dependence of our quoted results on the assumed scenario is,
we expect, small.
With a box size of $2h^{-1}{\rm\,Mpc}$ and a Lagrangian resolution
for the SLH code (Gnedin\markcite{G95} 1995; Gnedin \& 
Bertschinger\markcite{GB96} 1996) of $128^3$ which corresponds to a gas mass of
$\Delta M_g=3.2\times10^4 h^{-1} {\rm\,M}_{\sun}$ and a dark matter particle
mass of $\Delta M_d=3.7\times10^5 h^{-1} {\rm\,M}_{\sun}$, we have 
marginally resolved the smallest self-gravitating structures. Spatial 
resolution in our fiducial run is approximately $1h^{-1}\dim{kpc}$.

A detailed description of the physical modelling will be presented in
Gnedin \& Ostriker\markcite{GO96} (1996). Here we note  that we have allowed 
for the formation and destruction of molecular hydrogen, as well as
all other standard physical processes for a gas of primeval composition,
following in detail the ionization and recombination of all species in
the ambient radiation field. The spatially averaged but frequency dependent
radiation field, in turn, allows for sources
of radiation (quasars and massive stars), sinks (due to 
continuum opacities) and cosmological effects. In regions which are cooling
and collapsing we have allowed the formation of point-like ``stellar''
subunits, permitting them to release radiation and (in proportion)
metal rich gas, which we have considered in the treatment of cooling.
Finally, we have taken into account the fact that dense lumps will be
shielded from the background radiation field. This reduces the heating rates
for dense clumps and makes it nearly certain that once they have formed
and started to collapse, the process will be irreversible.

\section{A Qualitative Approach to Reionization}

Let us first ignore reheating and estimate the rate at which matter would 
begin to collapse following the standard theories of gravitational instability
such as the Press-Schechter picture\markcite{PS74} (1974). 
For cold gas in CDM-like theories
at first no mass elements are unstable to collapse. Then as gravitational
instability grows, higher and higher mass scales will become unstable, as
indicated by the line made of heavy dots in Fig.\ref{figMB}a.

\placefig{
\begin{figure}
\insertfigure{\figdir/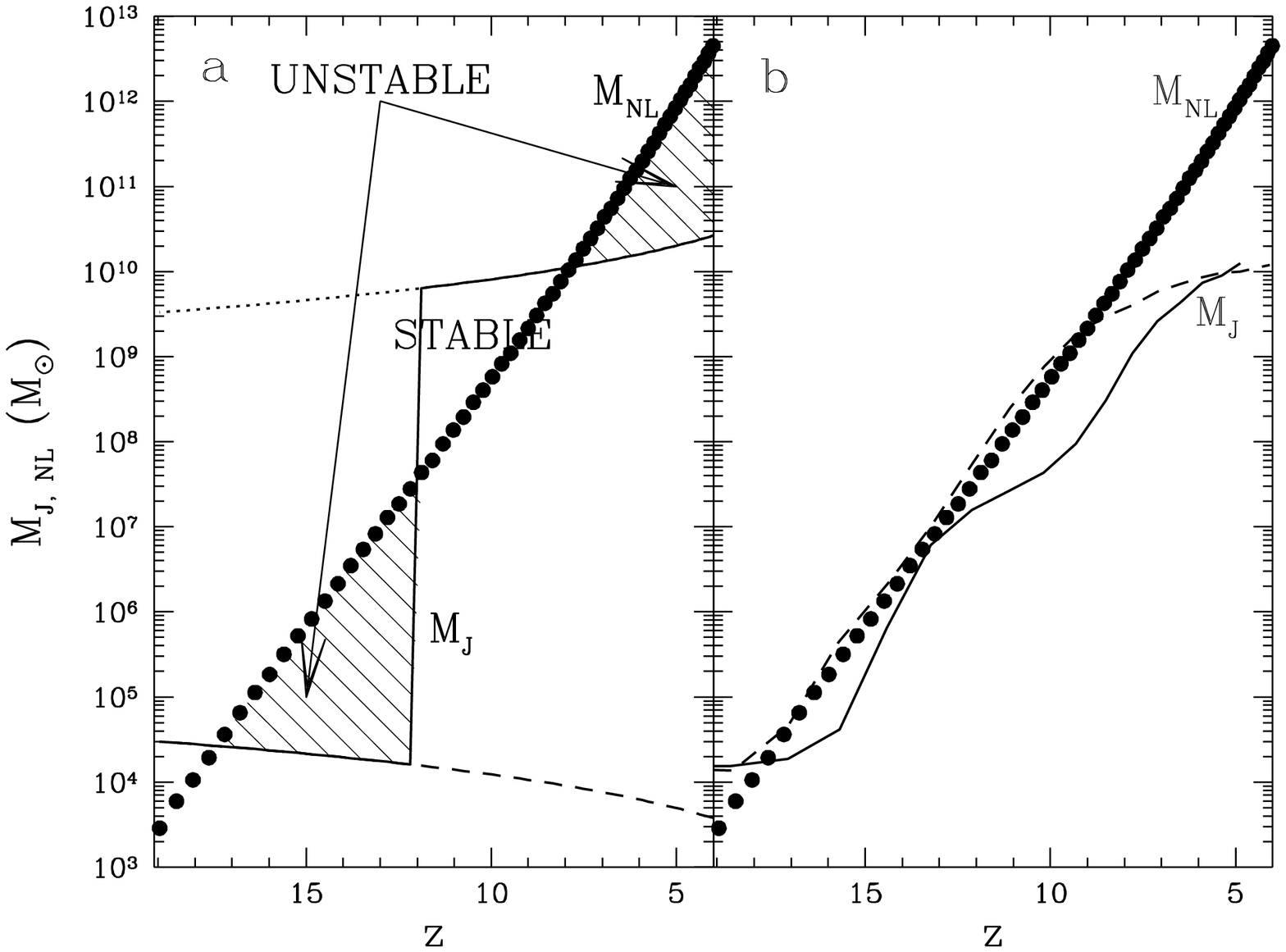}
\caption{\label{figMB}\capMB}
\end{figure}
}

Now consider the Jeans mass, the mass above which the self-gravitating
gas at the cosmic density and temperature would be unstable. The dashed curve
in Fig.\ref{figMB}a shows gas at the 
CBR temperature and the dotted curve marks
the gas at $10^4{\rm\,K}$, with a schematic transition curve indicated by the
solid curve for gas that heats rapidly to $10^4\dim{K}$ at $z=12$, when,
we hypothesize, ionization and reheating occur in this schematic
treatment.
We see that we might expect there to be an epoch of instability to exist before
reheating and reionization, when low mass objects collapse out of cold gas
(left hatched region) and another, later epoch, when galactic mass objects
condense out of $10^4{\rm\,K}$ gas (right hatched region), which are
separated in time by a region during which gas is stable against
gravitational collapse. A diagram of this type was
computed and shown in Cen, Ostriker, \& Peebles\markcite{COP93} 
(1993; Fig.8) to
illustrate, quantitatively, the epochs at which star and galaxy formation would
occur in the PBI model.

This early population of stars that formed before reheating
may be considered a candidate for a relatively smoothly distributed 
Population III. 
One expect that these stars will mostly reside in
low mass systems
($M_{\rm tot}\la10^{7.5}\dim{M}_{\sun}$, 
$M_{*}\la10^6\dim{M}_{\sun}$) which will contaminate
the universe with a first generation of metals (to $Z/Z_\odot\sim
10^{-4}$ by redshift of around 10).
After the Jeans mass increases significantly, normal 
galaxy formation would occur with masses $M_{g,\rm tot}>10^{10}\dim{M}_{\sun}$,
$M_{g,*}>10^8\dim{M}_{\sun}$ from gas already pre-enriched with metals
(Population II). However, as we shall see, while there are elements of 
truth to the scenario
indicated by Figure \ref{figMB}a, the real situation is far more complex.

\section{Results}

We have performed three different simulations (cf Table 1)
with different box sizes
\placefig{{\tableone}}
and resolution to assess the importance of different scales and estimate
the uncertainty due to the finite resolution of our simulations. We fix
cosmological parameters as follows: $\Omega_0=0.35$, $h=0.7$, and
$\Omega_b=0.03$. 
The largest of our simulations,
run A, is our fiducial run against which to compare results of 
the two smaller runs.
However, it happened by accident that the particular realization of initial
conditions we used for run A had substantially less power on large scales
($1-2h^{-1}{\rm\,Mpc}$) than the true power spectrum, and we have to take 
this fact into account when interpreting our results.

\subsection{Evolution of Jeans and Nonlinear Mass and Population III}

In the
real universe, unlike the oversimplified picture illustrated in 
Fig.\ref{figMB}a,
there are inhomogeneities on a range of scales, so
each small piece of the universe would evolve according to its local
values of the Jeans and nonlinear mass.
But
different pieces are not correlated, so we must appropriately average 
Fig.\ref{figMB}a over the
whole distribution of densities and temperatures in the universe to
compute the evolution of the average Jeans and nonlinear mass,
with the result being
that the two separate epochs of star formation would merge and the
Jeans mass, instead of jumping over the nonlinear mass in one sudden
reheating, would instead trace the nonlinear mass closely in a
self-regulating fashion. 

The Fig.\ref{figMB}b, our computed result, illustrates this
conclusion. In this figure we again mark with dots the nonlinear mass,
and Jeans mass from runs A (the solid curve) and B (the dashed curve)
are plotted on top of the nonlinear mass. We notice that they closely
follow each other from $z=17$ to $z=10$.
There are not 
two distinct epochs of star formation
separated by the period of reheating. Averaged globally, the Jeans mass, due 
to feedback, traces the evolution of the nonlinear mass scale.
Too high a rate of star formation produces too large a fraction of
the universe at $10^4\dim{K}$ with, consequently, too high an average
Jeans mass suppressing star formation -- and conversely.
But, as we shall see, there is a more complicated scenario, dependent on
molecular cooling, which does produce an early Pop III.

\subsection{Population II and Population III}

We now proceed to study individual objects formed in simulations. For all
bound objects identified with the DENMAX algorithm 
(Bertschinger \& Gelb\markcite{BG91} 1991) and containing more than 100 
particles
we compute their properties
like the total mass $M_{\rm tot}$, the total baryonic mass $M_{\rm b}$,
the mass in stars $M_{*}$, the mass of metal-enriched gas $M_Z$, the
average redshift of star formation $z_{\rm form}$ computed according
to the following formula $\log(1+z_{\rm form}) \equiv {\sum_* m_*\log(1+z_*)
/\sum_* m_*}$,
where the sum is over all star particles in the object, and $z_*$ is
the redshift of a star particle formation, and possibly other properties.

\placefig{
\begin{figure}
\insertfigure{\figdir/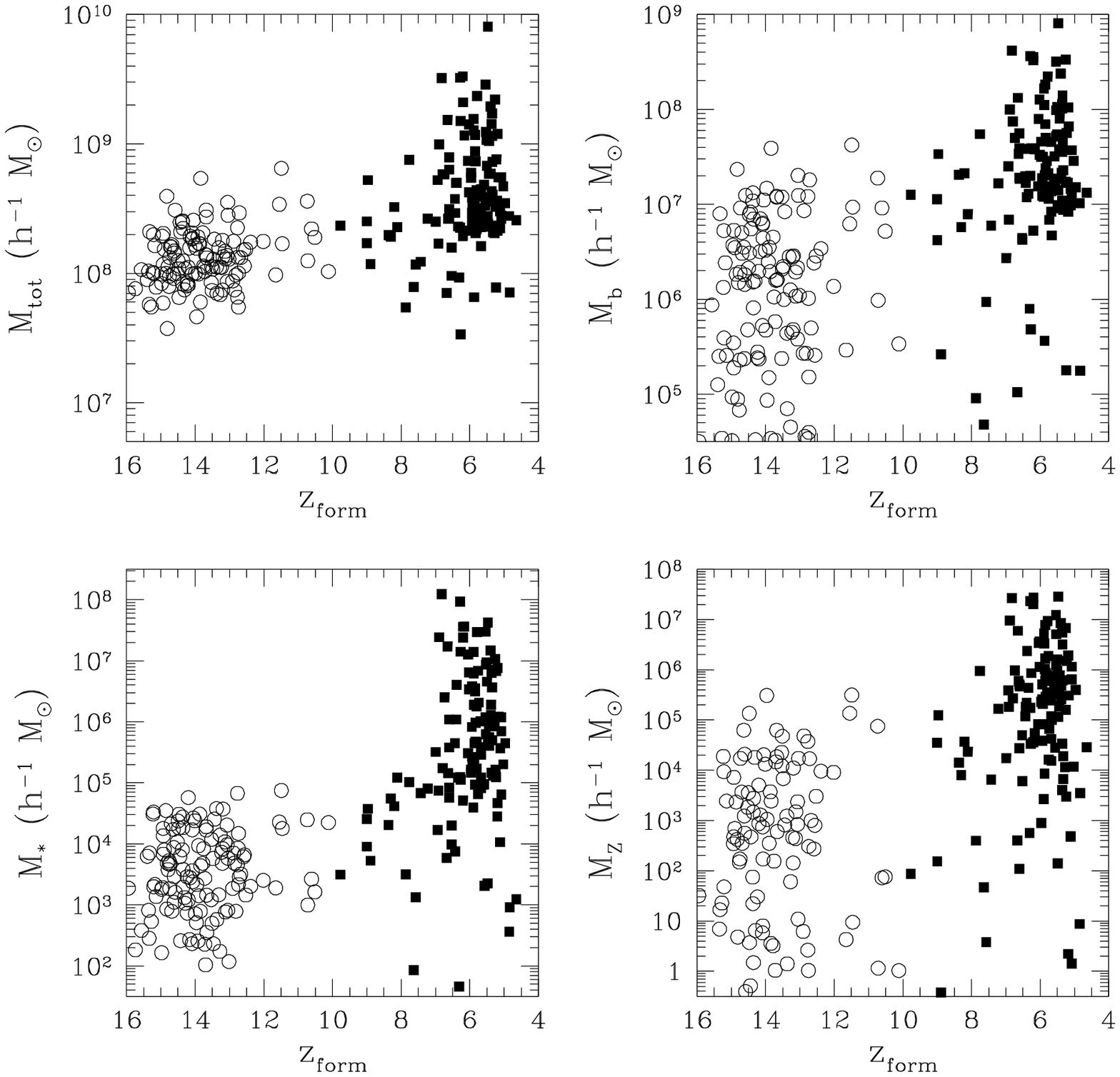}
\caption{\label{figZM}\capZM}
\end{figure}
}

We now plot in Fig.\ref{figZM} distributions of $M_{\rm tot}$, $M_{\rm b}$,
$M_*$, and $M_Z$ as a function of $z_{\rm form}$ for 
all bound object with more
than 100 particles identified at $z=4.4$ in run A. The distribution is
obviously bimodal, and to stress that, we plot all objects whose stars
formed at $z>10$ with open circles and all objects whose stars formed
after $z=10$ with solid squares. According to the KS test, the probability
that distributions of total masses for $z>10$ and $z<10$ objects are
drawn from the same random distribution is infinitesimal. Thus,
there indeed exist two populations of objects, which we
will call Population II ($z_{\rm form}>10$)
and Population III ($z_{\rm form}<10$), but not due to the evolution of the 
Jeans mass.
Distributions from runs B and C both show the same property, except that
the run B has only a handful of Pop III objects.

\placefig{
\begin{figure}
\epsscale{0.65}
\insertfigure{\figdir/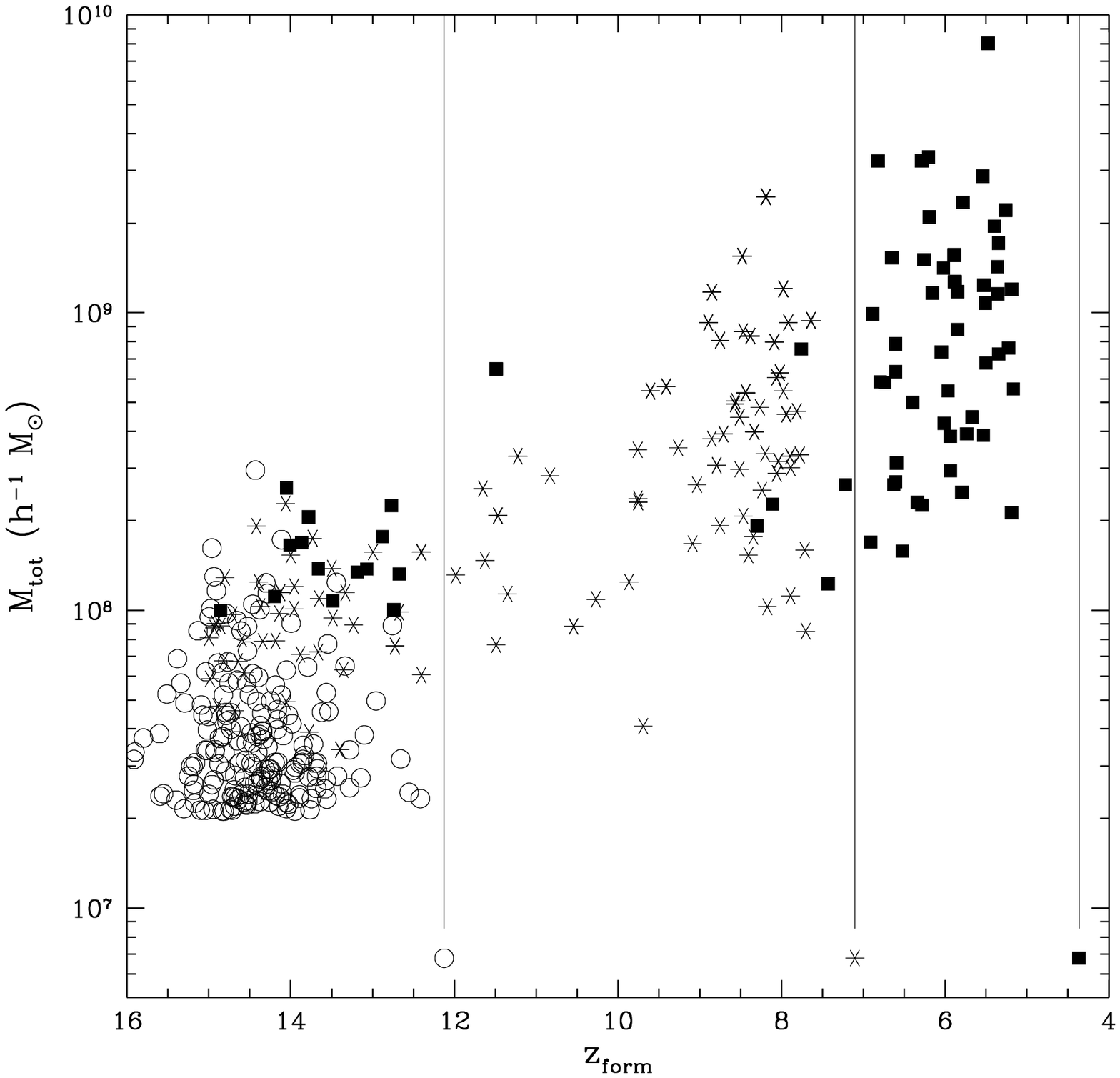}
\caption{\label{figZE}\capZE}
\end{figure}
}

In order to understand the appearance of two populations, we first concentrate
on their evolution. We identify all bound objects with more than 100 particles
at redshift $z=12.1$. Those are plotted as open circles in Fig.\ref{figZE}
and the vertical line with the open circle at the bottom shows the redshift
$z=12.1$. According to our criterion, they are all Pop III objects at that
time. We now track the evolution of those objects in 
($M_{\rm tot}$,$z_{\rm form}$)
plane and plot them with stars at $z=7.1$ and with solid squares at $z=4.3$.
It is apparent that with time Pop III objects are turning into Pop II.

This conclusion 
is not surprising if we recall that star formation is occurring in all objects,
and younger stars, formed after $z=10$, would drag the $z_{\rm form}$ label
to lower redshifts. However, this does not explain the presence of a gap
between Pop II and Pop III. If the star formation rate were continuous 
over the time,
there would be a continuous distribution of objects with respect to 
$z_{\rm form}$. Therefore, we must conclude that there existed two epochs
of star formation, the first initial burst at $z\sim14$, and the following
continuous star formation at $z<10$. Since the existence of those two epochs 
cannot be explained by the evolution of the Jeans mass, we must search for
another reason for it.

\subsection{Physical Mechanism for the Formation of Population III}

Radiative cooling of gas is responsible
for eventual gas collapse and star formation. 
\placefig{
\begin{figure}
\epsscale{0.65}
\insertfigure{\figdir/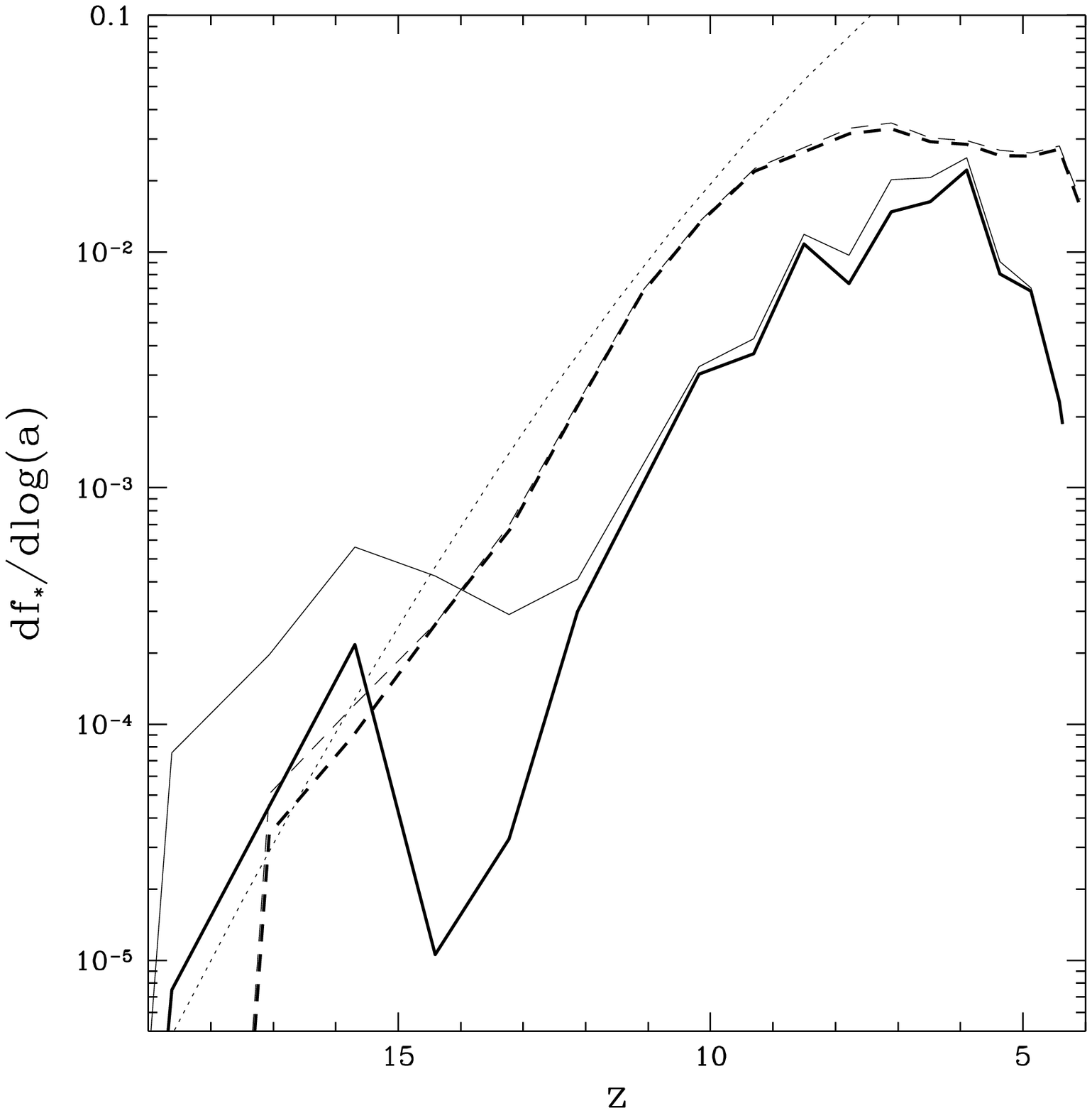}
\caption{
\label{figGF}\capGF}
\end{figure}
}
In Fig.\ref{figGF} we plot the star formation rates for run A (solid lines)
and run B (dashed lines). For each run we show as a heavy line the actual
star formation rate, and in the thin line the star formation rate that we would
compute assuming that the cooling time is infinitely short. For run B 
the difference is very small and only exists at $z=14-17$, whereas for run A 
there exist two distinct epochs of star formation which disappear if the
cooling time is set to zero. 
Also, shown with the thin dotted line is
the star formation rate computed from Press-Schechter formalism assuming
that all gas in halos with the virial temperature above $1.5\times10^4\dim{K}$
form stars.

\placefig{
\begin{figure}
\insertfigure{\figdir/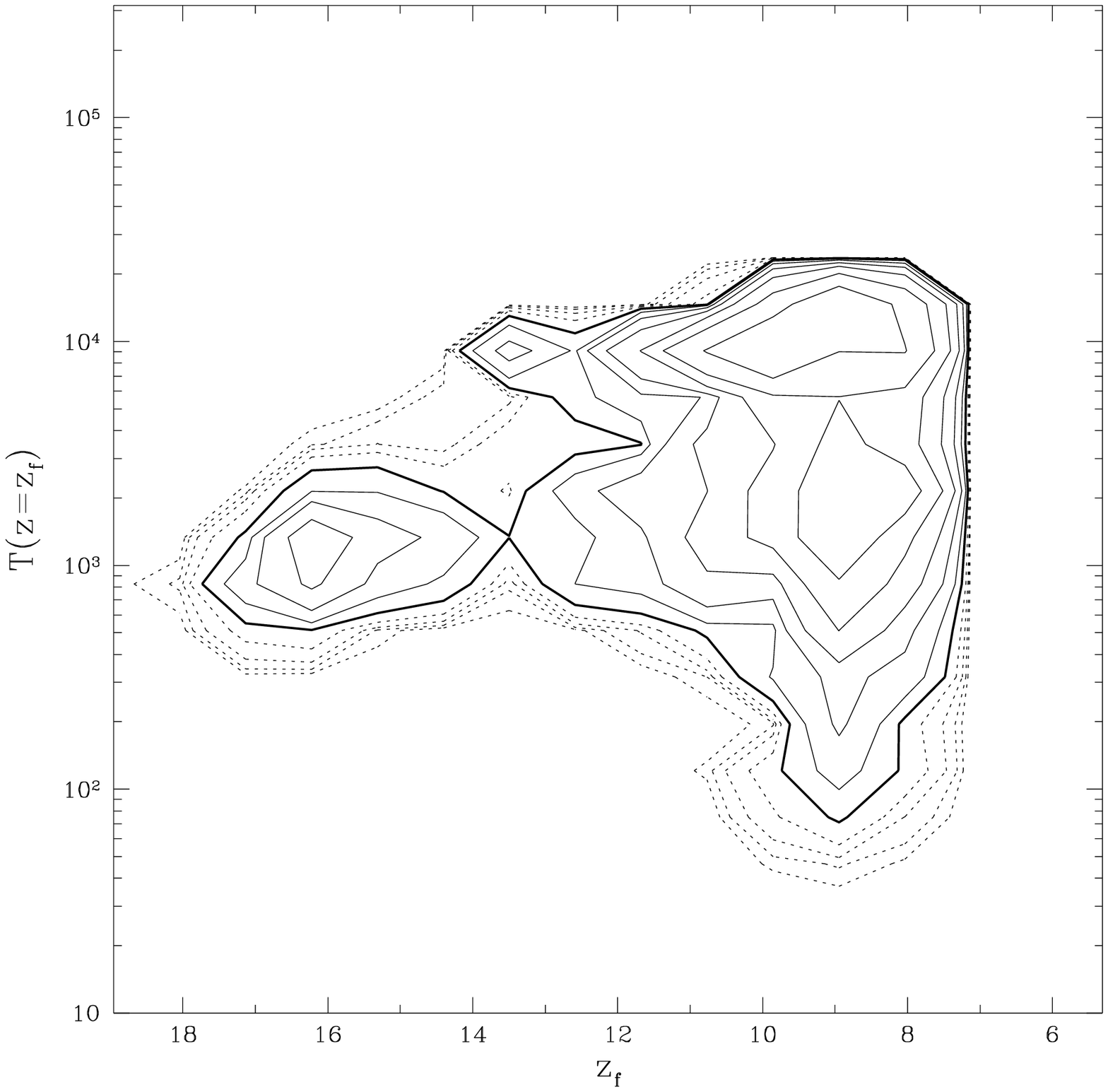}
\caption{\label{figTD}\capTD}
\end{figure}
}

By the time the first epoch of star formation is
completed, $z=14$, the density parameter in stars, $\Omega_*$ 
is reaching $10^{-5.5}$,
and the metal contamination is reaching $\bar Z/Z_{\sun}\sim10^{-3.7}$.

In order to understand the difference in cooling at high and low redshift
we compute the temperature and density for each star particle at the moment
it was created. Fig.\ref{figTD} shows joint distributions of all star particles
in [$T(z_{\rm form})$,$z_{\rm form}$]
plane for the fiducial run A.
Two epochs of star formation are easily observed in 
Fig.\ref{figTD}. The
careful examination of Fig.\ref{figTD} reveals that the
early peak corresponds to the temperatures around $10^3\dim{K}$. Since
the only cooling mechanism which exists at these temperatures in the plasma
of primeval composition is molecular hydrogen, we therefore conclude that
Population III stars formed at $z>14$ from primeval gas that collapsed 
losing it energy by exiting rotational and vibrational levels of hydrogen 
molecules, and Population II stars formed later from gas that collapsed,
losing its energy mainly by exiting hydrogen atomic lines. A detailed study
of individual objects confirms that early star formation is driven by $\MH$
cooling.

\placefig{
\begin{figure}
\insertfigure{\figdir/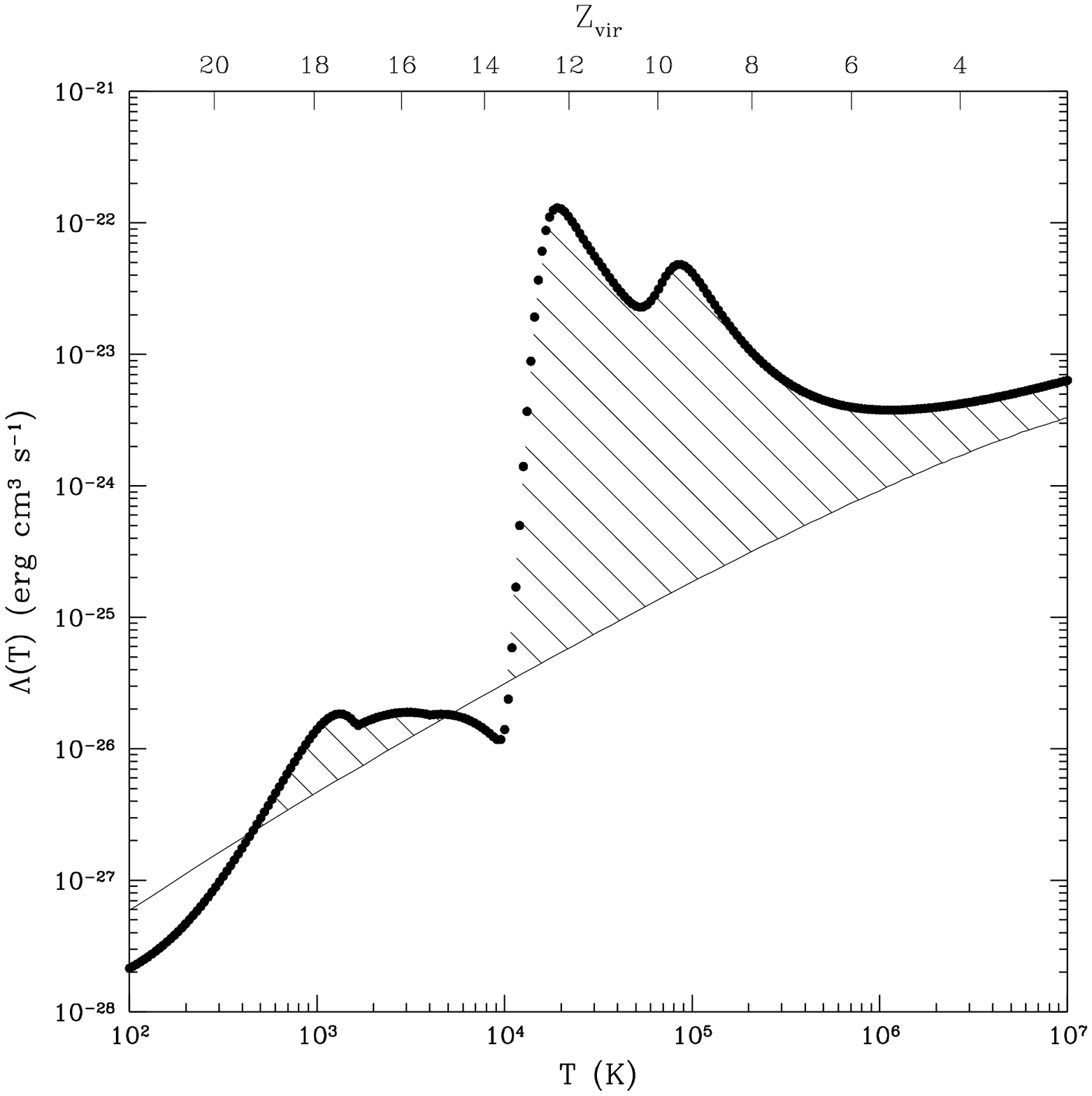}
\caption{
\label{figLL}\capLL}
\end{figure}
}

To illustrate this effect further, we show in Fig.\ref{figLL} with heavy
dots the cooling
function of the primeval gas containing $10^{-3}$ of molecular hydrogen my
mass at low temperature (at higher temperatures, $T>4,000\dim{K}$, the
destruction of molecular hydrogen is taken into account). The thin line
shows the value the cooling function should have in order for a virialized
object with given temperature to cool in a Hubble time. This figure can only
be considered as an illustration since it is based on the same oversimplified 
line of arguments as presented in \S 2. However, it indicates that
molecular hydrogen cooling is relatively more efficient at higher redshifts.

\section{Conclusions}

We have demonstrated by means of detailed cosmological hydrodynamic
simulations, which includes an accurate treatment of atomic physics
and chemistry of the primeval hydrogen-helium plasma, accounts approximately
for the self-shielding of the gas and cooling on metals, and incorporates
phenomenological approach to star formation, that the evolution of the Jeans
mass during reheating of the universe 
does not lead to formation of two distinct populations of
stars: Pop II and Pop III, but, instead, tracks the nonlinear mass due to feedback effects consequent to star formation.
Nevertheless, the two distinct populations
of stars form, for a completely different physical reason: the
first generation of stars formed when molecular hydrogen cooling
was important before $z=14$, and after molecular hydrogen was depleted
 by the rise of the virial
temperature in most of objects and photo-destruction, 
there was a relative pause in star formation until
sometime around $z=10$, when the fraction of objects with temperature in excess
of $10^4\dim{K}$ became significant and hydrogen line cooling became
the main cooling mechanism for Pop II star formation. 

At each particular epoch the most massive objects are statistically 
the oldest, they formed 
first, accreted more mass and became Pop II objects before 
less massive ones. Therefore, at every particular moment it is 
the Pop II objects that are oldest, and the time-sequence of formation
of Pop III and then Pop II objects is actually reversed, but it stems
merely from the fact that the Pop III identification is temporary, and
most Pop III objects will become Pop II objects with time.
Since the run B barely shows the existence of Pop III objects, its mass
resolution is apparently insufficient to resolve objects with the virial
temperature of around $10^3\dim{K}$, where the Pop III star formation occurs.
We conclude therefore that mass resolution of the order of
$10^4\dim{M}_{\sun}$ is required in order to find Pop III objects in
a simulation, and that in the real world it is likely that this population
did form at redshift $z>14$ leading to an early contamination of the universe
with metals $\bar Z/Z_{\sun}\sim10^{-3.7}$.

\acknowledgements

The authors would like to express their gratitude to Prof.\ Martin Rees for
numerous fruitful discussions and valuable comments.
This work was supported by NSF grant AST-9318185
awarded to the Grand Challenge Cosmology Consortium.

\placefig{\end{document}}

\clearpage

\tableone

\clearpage
\newcounter{figurecap}
\setcounter{figurecap}{0}

\begin{center}
\bf Figure Captions
\end{center}

\refstepcounter{figurecap}
Fig.\thefigurecap---\label{figMB}\capMB

\refstepcounter{figurecap}
Fig.\thefigurecap---\label{figZM}\capZM

\refstepcounter{figurecap}
Fig.\thefigurecap---\label{figZE}\capZE

\refstepcounter{figurecap}
Fig.\thefigurecap---\label{figGF}\capGF

\refstepcounter{figurecap}
Fig.\thefigurecap---\label{figTD}\capTD

\refstepcounter{figurecap}
Fig.\thefigurecap---\label{figLL}\capLL

\end{document}